\documentclass[letterpaper,aps,prl,reprint,showpacs,superscriptaddress,amsmath,amssymb]{revtex4-1}

\usepackage{graphicx}
\usepackage[latin1]{inputenc}
\usepackage{dcolumn}
\usepackage{hyperref}

\begin{document}

\newcommand*{\mean}[1]{\left\langle #1 \right\rangle}
\newcommand*{\ket}[1]{\left|#1\right\rangle}
\newcommand*{\bra}[1]{\left\langle #1\right|}
\newcommand*{\abs}[1]{\left|#1\right|}
\newcommand*{\tr}{\mathrm{tr}}
\newcommand*{\bo}[1]{\mathbf{#1}}
\newcommand*{\norm}[1]{\left\Vert #1 \right\Vert }
\newcommand*{\DPZ}[3]{\left.\frac{\partial#1}{\partial#2}\right|_{#3}}
\renewcommand*{\i}{\mathrm{i}}

\title{Ultimate sensitivity of precision measurements with Gaussian quantum 
light : \\a multi-modal approach}

\author{Olivier Pinel}
\affiliation{Laboratoire Kastler Brossel,  Université Pierre et Marie Curie-Paris 6,\\
ENS, CNRS; 4 place Jussieu, 75252 Paris, France}
\author{Julien Fade}
\affiliation{Institut Fresnel, CNRS, Aix-Marseille Université, \\
\'Ecole Centrale Marseille, Campus de Saint-Jérôme, 13013 Marseille, France}
\author{Daniel Braun}
\affiliation{Laboratoire de Physique Théorique, Université Paul Sabatier, Toulouse III, \\
118 route de Narbonne 31062 Toulouse, France}
\author{Pu Jian}
\affiliation{Laboratoire Kastler Brossel,  Université Pierre et Marie Curie-Paris 6,\\
ENS, CNRS; 4 place Jussieu, 75252 Paris, France}
\author{Nicolas Treps}
\affiliation{Laboratoire Kastler Brossel,  Université Pierre et Marie Curie-Paris 6,\\
ENS, CNRS; 4 place Jussieu, 75252 Paris, France}
\author{Claude Fabre}
\affiliation{Laboratoire Kastler Brossel,  Université Pierre et Marie Curie-Paris 6,\\
ENS, CNRS; 4 place Jussieu, 75252 Paris, France}
\date{\today}

\begin{abstract}

Multimode Gaussian quantum light, which includes multimode squeezed and multipartite quadrature entangled light, is a very general and powerful quantum resource with promising applications in quantum information processing and metrology. In this paper, we determine the ultimate sensitivity in the estimation of any parameter when the information about this parameter is encoded in such light, irrespective of the information extraction protocol used in the estimation and of the measured observable. In addition we show that an appropriate homodyne detection scheme allows us to reach this ultimate sensitivity. We show that, for a given set of available quantum resources, the most economical way to maximize the sensitivity is to put the most squeezed state available in a well-defined light mode. This implies that it is not possible to take advantage of the existence of squeezed fluctuations in other modes, nor of quantum correlations and entanglement between different modes. 
\end{abstract}

\pacs{03.65.Ta, 42.50.Ex, 42.50.Lc, 42.50.St}

\maketitle

Optical techniques are widely used in many areas of science and technology to make accurate measurements and diagnostics, from microscopy, spectrography, chemical analysis, to gravitational wave detection and ranging. There are many reasons for this: light allows us to extract information in a remote and non destructive way, it carries information in a massively parallel way, and perhaps more importantly, optical measurements can reach very high precision and sensitivity levels.

It is therefore important to know what is the ultimate limit of sensitivity that can be possibly achieved in the estimation of a parameter $\theta$ that is encoded by one way or another in a light beam, given some constraints such as a fixed mean photon number $N$. This limit is imposed by the unavoidable quantum fluctuations of light and depends on the quantum state of light which conveys the information about $\theta$. When the light is in a coherent state, this limit is called  `standard quantum limit' and scales as $1/N^{1/2}$.

Many studies have been devoted to finding ways to enhance the sensitivity of parameter estimation beyond the standard quantum limit using quantum resources. It has been shown that enhanced sensitivity can be achieved by using squeezed light \cite{Caves81} or entangled light \cite{Giovanetti06}. This has been first experimentally demonstrated for measurements in which the information about the parameter $\theta$ is carried by the total intensity \cite{Marin97} or by the phase \cite{Xiao87} of a light beam. Later situations were considered where the parameter $\theta$ does not change the total intensity of the light but modifies the details of the repartition of light in the transverse plane \cite{Delaubert08} (for example to estimate a very small lateral displacement of a beam \cite{Treps03}). As the energy of the squeezed state increases with the squeezing factor, the ultimate limit with squeezed state for a fixed total energy scales as $1/N^{3/4}$.

If one uses instead entangled states such as NOON states \cite{Kok04} one reaches the so-called Heisenberg-limit (HL) which scales as $1/N$. However, in the present state of technology real measurement schemes using these states do not lead to very high sensitivities, because of the small values of $N$ experimentally reachable (so far, the highest achievable NOON state has $N\sim100$ \cite{Higgins07}), and decoherence tends to rapidly destroy these states, therefore limiting the performance of the measurement to a $1/N^{1/2}$ scaling for large $N$ \cite{Huelga97,Kolodynski10,Escher11}. In \cite{Braun11} a scheme was proposed that reaches the HL without the use of
an entangled state.

\begin{figure}[htbp]
\includegraphics[width=8cm]{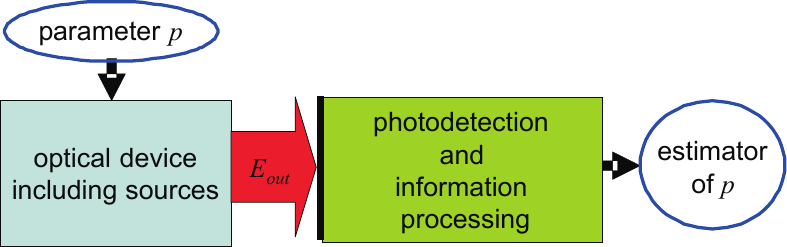}
\caption{General scheme for estimating light parameters.}
\label{setup}
\end{figure}

This paper tackles the problem of optimized parameter estimation in
a more practical way. Light is considered as a probe to measure a parameter of a physical system (see Fig.~\ref{setup}). As in that case all quantum limits scale as some inverse power on $N$, it is very important to consider states with very high $N$ values. It turns out that, so far, only multimode Gaussian states are the available non classical states of light with very high mean photon number. They include quantum resources like multimode squeezing and multipartite entanglement that are widely used in quantum optics and quantum information processing. These states are already generated experimentally with impressive amounts of squeezing \cite{Mehmet10} and entanglement \cite{Laurat05} shared by many modes \cite{Yukawa08}. When they include a coherent state in one of the modes, the mean photon number $N$  can be easily as large as $10^{16}$\cite{Keller}. 

The originality of the present approach is its multi-modal character. A multimode quantum state is defined not only by the value of the coefficients of its decomposition on the multimode Fock state basis $|n_1, n_2,...,n_{\ell} \rangle$ but also on the spatio-temporal shape of the different modes on which these Fock states are defined. This leaves us two kinds of degrees of freedom on which to act: as we will see below, the ultimate sensitivity is obtained not only by choosing the best possible Gaussian quantum state, but also by putting this state in an optimized mode basis.

\textit{Expression of the Quantum Cramer Rao bound for pure states}  - Our aim is to measure the smallest possible variation of a parameter $\theta$ around a given value that we take to be $0$ by an appropriate change of origin.  The quantum state which contains the information about this parameter  is described by a density matrix $\hat{\rho}_{\theta}$. The error in the estimation of $\theta$ based on $Q$ repeated measurements of an observable $\hat{A}$ on this state is given by \cite{Braunstein94}
\begin{equation}
\delta\theta=\frac{\mean{\delta A_{\mathrm{est}}^{2}}_{\theta}^{1/2}}{\sqrt{Q}\abs{\frac{\partial}{\partial\theta}\mean{A_{\mathrm{est}}}_{\theta}}},
\end{equation}
where $A_{\mathrm{est}}$ is an unbiased estimator of $\theta$ that depends on the results of the measurements of $\hat{A}$. By optimizing over all estimators $A_{\mathrm{est}}$ and all measurements, Braunstein and Caves \cite{Braunstein94} showed that the best achievable sensitivity for measuring a small variation of $\theta$ is bounded by the so-called quantum Cramér-Rao (QCR) bound
\begin{equation}
\delta\theta\geq\delta\theta_{\mathrm{min}}\equiv \left( 2\sqrt{Q}\frac{s(\hat{\rho}_{\theta},\hat{\rho}_{\theta+d\theta})}{d\theta}\right)^{-1},
\end{equation}
where $s(\hat{\rho}_{\theta},\hat{\rho}_{\theta+d\theta})$ is the Bures distance between $\hat{\rho}_{\theta}$ and $\hat{\rho}_{\theta+d\theta}$, which, in the case of pure states $\ket{\psi_{1}}$ and $\ket{\psi_{2}}$ is equal to $\sqrt{2(1-\left|\mean{\psi_{1}|\psi_{2}}\right|)}$.

Let us now consider a pure quantum state of light
$\ket{\psi_{\theta}}$ spanning over $M$ different spatial or temporal
modes $\{v_{i}(\bo r,t)\}$ ($i=1,...,M$). For mixed states with
parameter independent mixing probabilities, the sensitivity can at
most be as good as for the pure states from which 
it is mixed \cite{Braun10}. We call $\hat{a}_{i}$ the annihilation operator in the mode $v_{i}$, and introduce the quadrature operators $\hat{x}_{i}=\hat{a}_{i}+\hat{a}_{i}^{\dagger}$ and $\hat{p}_{i}=\i(\hat{a}_{i}^{\dagger}-\hat{a}_{i})$. We define the column vectors $\hat{\bo x}=(\hat{x}_{1},\ldots,\hat{x}_{M})^{\top}$, $\hat{\bo p}=(\hat{p}_{1},\ldots,\hat{p}_{M})^{\top}$, and $\hat{\bo X}=(\hat{\bo x},\hat{\bo p})^{\top}$.  
	
The overlap between the states $\ket{\psi_{\theta}}$ and $\ket{\psi_{\theta+d\theta}}$ reads:
\begin{equation}
\left|\mean{\psi_{\theta}|\psi_{\theta+d\theta}}\right|^{2}=\left(4\pi\right)^{M}\int W_{\theta}(\bo X)W_{\theta+d\theta}(\bo X)\ d^{2M}\!\bo X ,
\end{equation}
$W_\theta$ being the Wigner function of $\ket{\psi_\theta}$:
\begin{equation}
W_\theta(\bo x,\bo p)=\frac{1}{(2\pi)^{M}}\int e^{i\boldsymbol{\xi}.\bo p} \langle \bo x-\boldsymbol{\xi} \ket{ \psi_\theta} \bra{\psi_\theta}\bo x+\boldsymbol{\xi}\rangle \ d^{M}\!\boldsymbol{\xi}
\end{equation}
 At second order in $d\theta$, the overlap 
is equal to
\begin{equation}
\left|\mean{\psi_{\theta}|\psi_{\theta+d\theta}}\right|^{2}\simeq1-\frac{d\theta^{2}}{2}\left(\left(4\pi\right)^{M}\int\left(W_{\theta}^{\prime}(\bo X)\right)^{2}\ d^{2M}\!\bo X\right).
\end{equation}
The first order vanishes because the states are pure. Throughout this letter, for any function depending on the parameter $\theta$, we use the convention $f_{\theta}^{\prime}\equiv\DPZ f{\theta}{\theta=0}$, regardless of what other explicit variables $f$ might depend on. 

This leads to the QCR bound for pure states
\begin{equation}
\delta\theta_{\mathrm{min}}=\left( 2 Q \left(4\pi\right)^{M}\int\left(W_{\theta}'(\bo X)\right)^{2}\ d^{2M}\!\bo X \right)^{-1/2}\label{eq:QCR_wigner} .
\end{equation}
This intermediate result is very interesting as it gives a simple expression of the QCR bound for any pure quantum state. In the remainder of this paper, we will apply this formula to Gaussian states.

\textit{QCR bound for pure Gaussian states} - For a Gaussian state $\ket{\psi_{\theta}}$, the Wigner function takes the form 
\begin{equation}
W_{\theta}(\bo X)=\frac{1}{(2\pi)^{M}}\exp\left(-\frac{1}{2}(\bo X-\overline{\bo X}_{\theta})^{\top}\boldsymbol{\Gamma}_{\theta}^{-1}(\bo X-\overline{\bo X}_{\theta})\right)
\end{equation}
where $\overline{\bo X}_{\theta}$ is the column vector of the expectation values of the quadratures for the different modes, and  $\boldsymbol{\Gamma}_{\theta}$  the symmetrized covariance matrix: $\boldsymbol\Gamma_{\theta,[i,j]}= \frac{1}{2} (\mean{\bo X_i \bo X_j}_\theta + \mean{\bo X_j \bo X_i}_\theta)$.  As we treat the problem in all its generality, both possibly depend on $\theta$.  One finds from (\ref{eq:QCR_wigner}) 
\begin{equation}
\delta\theta_{\mathrm{min}}=Q^{-1/2} \left(\overline{\bo X}_{\theta}^{\prime\top}\boldsymbol{\Gamma}_{\theta}^{-1}\overline{\bo X}_{\theta}^{\prime}+\frac{\tr\left(\left(\boldsymbol{\Gamma}_{\theta}^{\prime}\boldsymbol{\Gamma}_{\theta}^{-1} \right)^{2}\right)}{4}\right)^{-1/2} \label{eq:QCRGauss}.
\end{equation}
The expression in the outermost 
bracket of Eq.~(\ref{eq:QCRGauss}) corresponds to the quantum Fisher
information $I_\mathrm{Fisher}$ for a pure Gaussian state. It consists
of two terms which represent the information about $\theta$ that can be extracted respectively from the mean field and from the noise.  In the limit of very large values of $N$, more precisely when the quantum field fluctuations are so small compared to $N$ that one can treat them to first order, the second term turns out to be negligible compared to the first, and we will neglect it from now on. This approximation is a consequence of the practical approach we consider in this paper and corresponds to realistic experimental implementations. Let us stress that such a linearization procedure has been widely used in the literature to determine the Gaussian quantum state which is produced by nonlinear effects such as parametric down-conversion or four-wave mixing.

Let us now use our freedom of choice of the mode basis in which to describe the quantum state: we will see that $I_\mathrm{Fisher}$ can be expressed in more physical terms if one introduces a mode basis $\{ \widetilde{v}_{i}(\bo r,t) \}$ specific to our problem. We first define the \textit{normalized mean photon field mode} as: 
\begin{equation}\label{fieldmode}
u_{\theta}(\bo r,t)=\frac{\overline{a}_{\theta}(\bo r,t)}{\norm{\overline{a}_{\theta}}}
\end{equation}
where $\hat{a}(\bo r,t)=\sum_{i}\hat{a}_{i}v_{i}(\bo r,t)$  is the local annihilation operator, $\overline{a}_{\theta}(\bo r,t)=\bra{\psi_{\theta}}\hat{a}(\bo r,t)\ket{\psi_{\theta}}$ the mean photon field, and $\norm{\overline{a}_{\theta}}$ its norm:
\begin{equation}
\norm{\overline{a}_{\theta}}=\left( \int |\overline{a}_{\theta}(\bo r,t)|^2 \ d^2\bo r dt \right)^{1/2} ,
\end {equation}
where the spatial integration is made over a surface perpendicular to the light beam propagation, and the time integration over the detection time. In the limit of a narrow-band field, the mean photon field mode $u_{\theta}$ is proportional to the mean value of the electric field in the $\theta$-dependent quantum state.

We can now define the \textit{detection mode} by
\begin{equation}\label{detmode}
\widetilde{v}_{1}(\bo r,t)=\frac{\overline{a}_{\theta}^{\prime}(\bo r,t)}{\norm{\overline{a}_{\theta}^{\prime}}} .
\end{equation}
One then completes the basis starting with mode $\widetilde{v}_1$ by other orthonormal modes $\widetilde{v}_{n>1}$. The modes $\widetilde{v}_n$ do not depend on $\theta$ since the derivative in (\ref{detmode}) has been taken at the value $\theta=0$

The expression of the Fisher information in the $\{ \widetilde{v}_{i}(\bo r,t) \}$ mode basis is very simple as it involves only one matrix element of $\boldsymbol{\Gamma}_{\theta}^{-1}$:
\begin{equation}
I_\mathrm{Fisher}= 4 \boldsymbol{\Gamma}_{\theta=0,\left[1,1\right]}^{-1} \norm{\overline{a}_{\theta}^{\prime}}^2
\end{equation}
where $\boldsymbol{\Gamma}_{\theta=0,\left[1,1\right]}^{-1}$ is the first left, top element of the matrix $\boldsymbol{\Gamma}_{\theta}^{-1}$ in the basis $\{ \widetilde{v}_{i}(\bo r,t) \}$ taken at the value $\theta=0$ of the parameter.

In particular, the Fisher information for a single measurement involving a coherent state ($\boldsymbol{\Gamma}_{\theta}=\boldsymbol{1}$), that we will call $I_0$, is found to be
\begin{equation}
I_0 = 4\norm{\overline{a}_{\theta}^{\prime}}^{2}=N_{\theta}\left(4 \norm{u_{\theta}^{\prime}}^{2}+\left(\frac{N_{\theta}^{\prime}}{N_{\theta}}\right)^{2}\right) ,
\end{equation}
where $N_{\theta}=\norm{\overline{a}_{\theta}(\bo r,t)}^2$ is a quantity that tends to the mean photon number $N$ in the high $N$ limit where fluctuations can be linearized. We obtain finally the following expression of the QCR bound for parameter estimation using quantum Gaussian states
\begin{equation} \label{QCRfinal}
\delta \theta_{\mathrm{min}}=\left[ QN_{\theta}\left(4\norm{u_{\theta}^{\prime}}^{2}+\left(\frac{N_{\theta}^{\prime}}{N_{\theta}}\right)^{2}\right)\boldsymbol{\Gamma}_{\theta,\left[1,1\right]}^{-1} \right]^{-1/2} .
\end{equation}
It depends on $3$ factors: the first one is as usual the mean total number of photons measured $QN_{\theta}$. The second one is related to the variation as a function of $\theta$ of the displacement of the mean field mode and the mean photon number. The more the light properties are affected by the variation of $\theta$, the better the sensitivity one can expect for its estimation. While the general argument is obvious, the explicit formula (\ref{QCRfinal}) is not. The last factor is the influence on the measurement of the quantum fluctuations of the state, which is remarkably contained in a single element of the inverse covariance matrix in our specific mode basis.

\textit{Optimized multimode Gaussian state for parameter estimation} - Let us now discuss under which conditions nonclassical multimode Gaussian states  can be put to best use in the estimation of $\theta$. We will take the point of view of an experimentalist who wants to use the minimum possible amount of quantum resources that allow him to reach the QCR bound. He will start from the simplest way known to date to generate multimode quantum Gaussian states \cite{Braunstein05}, which consists in linearly mixing several single mode squeezed beams produced by independent "squeezers", such as degenerate parametric amplifiers. We will call $\sigma_{\min}^{2}$ the smallest quadrature noise among all these squeezed modes. $\sigma_{\min}^{-2}$ is the largest eigenvalue of the inverse covariance matrix in the initial basis of the independent squeezed modes.  With the help of linear couplers i.e.~of a $\theta$ dependent unitary transformation of the mode basis, the multimode squeezing can be transformed into a multimode entangled/squeezed Gaussian state in a mode basis the spatio-temporal shape of which can also be tailored at will \cite{Morizur:2011jb}. One can show that, under such unitary transformations, the diagonal matrix elements of the inverse of the covariance matrix are bound by the spectral radius of $\boldsymbol{\Gamma}^{-1}_{\theta=0}$, which is equal to $1/\sigma_{\text{min}}^2$. Equality is reached only if the detection mode 1 is an eigenmode of the covariance matrix with the eigenvalue $\sigma_{\text{min}}^2$, and thus \emph{when the most squeezed state is put in the detection mode, with no quantum correlations with any other mode}. The QCR bound corresponding to the quantum resources that we have just described is thus
\begin{equation} \label{QCRopt}
\delta \theta_{\mathrm{min}}=\frac{\sigma_{\text{min}}}{\sqrt{QN_{\theta}}} \left(4\norm{u_{\theta}^{\prime}}^{2}+\left(\frac{N_{\theta}^{\prime}}{N_{\theta}}\right)^{2}\right)^{-1/2} .
\end{equation}

We have shown here an important result: the only way to saturate the Cramér-Rao bound in the configuration that we have just described is to put the most squeezed state available into the detection mode and not to have correlations with the other modes. The presence of other squeezed modes, or of any kind of entanglement, will not help to improve the sensitivity: one cannot take advantage of squeezed fluctuations or quantum correlations coming from different modes to improve the estimation of a single parameter \cite{Tilma10}. We therefore advise experimentalists to produce a single vacuum squeezed state, to put it in the detection mode, and to mix it with a coherent state of high mean photon number $N$ in the mean photon field mode $u_{\theta}(\bo r,t)$ . Doing that, they will be sure that nobody else will make a more sensitive estimation of the variation of $\theta$ around $0$ for a given shape $u_{\theta}(\bo r,t)$ of the mean field.

\textit{A possible experimental implementation that reaches the QCR bound} - The determination of the Quantum Cramér-Rao bound is very general and does not tell us which kind of detection, and which kind of measurement strategy is to be used in order to reach it. We show in this paragraph that a homodyne detection scheme in which the local oscillator is precisely taken in the detection mode allows us to reach the QCR bound.

 If one uses an intense local oscillator in mode $\widetilde{v}_{1}$, the balanced homodyne detection operator, for a null relative phase between the local oscillator and the measured beam,  is given by $\hat{D}={\hat {\tilde x}}_{1} \sqrt{N_{\mathrm{LO}}} $, where $N_{\mathrm{LO}}$ is the mean photon number of the local oscillator and ${\hat {\tilde x}}_{1}$ the real quadrature operator of the mode  $\widetilde{v}_1$. A balanced detection set-up therefore allows us to measure the projection of a multimode field on the oscillator mode, even in presence of many other modes.

For a small variation of the parameter $\theta$ around $0$ the mean value of the homodyne signal is given by:
\begin{eqnarray}
\mean{\hat D}_{\theta}&=& \sqrt{N_{LO}} \mean{\hat {\tilde x}_{1}}_{\theta}\nonumber \\
&=& 2 \sqrt{N_{LO}} \, {\mathcal Re}\left(\int \widetilde{v}_{1}^{*}\overline{a}_{\theta} \ d^2\bo r dt\right)
\end{eqnarray}
using the orthonormality properties of the mode basis $\{ \widetilde{v}_{i}(\bo r,t) \}$.  As 
\begin{equation}
\overline{a}_{\theta}\approx\overline{a}_{\theta=0}+\theta \, \overline{a}_{\theta}^{\prime} ,
\end{equation}
one finally gets by using the orthonormality properties of the mode basis $\{\widetilde{v}_{i}(\bo r,t) \}$, and the fact that $\int u_{\theta}^{*}u_{\theta}^{\prime}\ d^2\bo r dt$ is a purely imaginary number,
\begin{equation}\label{homo1}
\mean{\hat D}_{\theta}=\sqrt{N_{LO}} \left( \sqrt{I_{0}} \theta+2\frac{N_{\theta}^{\prime}}{\sqrt{I_{0}}}\right).
\end{equation}
The homodyne signal, suitably calibrated, is therefore an estimator of $\theta$. Because of the additional term in (\ref{homo1}), the estimation is biased. We then introduce the unbiased estimator $\widetilde{\theta}$ of $\theta$,
\begin{equation}
\widetilde{\theta}= \frac{\mean{\hat{D}}_{\theta}-D_{0}}{\sqrt{N_{\mathrm{LO}}I_{0}}}.
\end{equation}
where  $D_{0}$  is the mean value of $\hat{D}$ for $\theta=0$.  Considering the case when the light state is squeezed in the detection mode by a factor $\sigma_\mathrm{min}^{2}$ and assuming a unity signal to noise ratio, the sensitivity of the homodyne measurement can be shown to be:
\begin{equation}
\delta\theta_\mathrm{homodyne}=\frac{\sigma_\mathrm{min}}{\sqrt{N_{\theta}\left(4\norm{u_{\theta}^{\prime}}^{2}+\left(N_{\theta}^{\prime}/N_{\theta}\right)^{2}\right)}}
\end{equation}
which is indeed equal to the QCR bound (\ref{QCRopt}) for a single measurement.

In conclusion, we have derived the expression of the ultimate limit for parameter estimation using pure Gaussian multimode states. We have shown that this limit can be reached with the help of a balanced homodyne detection scheme.  We have also shown that multimode squeezing and multipartite entanglement are of no help, and that it is very important to shape in the best way the mode in which to put the non-classical Gaussian state in order to reach the ultimate limit in the most economical way. These results are good news for the experimentalists because single mode highly squeezed Gaussian states can be readily generated experimentally and because a simple homodyne detection scheme, easily achievable in a laboratory, is sufficient for reaching the best possible sensitivity.

\begin{acknowledgments}
We acknowledge the financial support of the Future and Emerging Technologies (FET) programme within the Seventh Framework Programme for Research of the European Commission, under the FET-Open grant agreement HIDEAS, number FP7-ICT-221906, and of the ANR project QUALITIME.
\end{acknowledgments}

\end{document}